\documentclass[%
reprint,
showpacs,
superscriptaddress,
longbibliography,
amsmath,amssymb,
aip,
author-year,
]{revtex4-1}

\RequirePackage{graphicx}
\RequirePackage{amssymb}

\usepackage{xspace}														
\newcommand{\degree}{$^\circ$\xspace}                					
\newcommand{\PFTOm}{$P\bar{4}2_1m$\xspace}          					
\newcommand{\bcgo}{Ba$_2$CoGe$_2$O$_7$\xspace}          				
\newcommand{\ccso}{Ca$_2$CoSi$_2$O$_7$\xspace}         					
\newcommand{\scso}{Sr$_2$CoSi$_2$O$_7$\xspace}          				
\newcommand{\bmgo}{Ba$_2$MnGe$_2$O$_7$\xspace}          				

\begin{document}                  


\title{Origin of forbidden reflections in multiferroic Ba$_2$CoGe$_2$O$_7$ by neutron diffraction: Symmetry lowering or Renninger effect?\vspace*{3ex}}

\author{Andrew Sazonov}
\email{mail@sazonov.org}
\affiliation{Institute of Crystallography, RWTH Aachen University and J\"ulich Centre for Neutron Science (JCNS) at Heinz Maier-Leibnitz Zentrum (MLZ), 85747 Garching, Germany}
\author{Martin Meven}
\affiliation{Institute of Crystallography, RWTH Aachen University and J\"ulich Centre for Neutron Science (JCNS) at Heinz Maier-Leibnitz Zentrum (MLZ), 85747 Garching, Germany}
\author{Georg Roth}
\affiliation{Institute of Crystallography, RWTH Aachen University, 52056 Aachen, Germany}
\author{Robert Georgii}
\affiliation{Heinz Maier-Leibnitz Zentrum (MLZ) and Physics Department E21, Technical University of Munich, 85747 Garching, Germany}
\author{Istv\'an K\'ezsm\'arki}
\affiliation{Department of Physics, Budapest University of Technology and MTA-BME Lend\"ulet Magneto-optical Spectroscopy Research Group, 1111 Budapest, Hungary}
\author{Vilmos Kocsis}
\affiliation{Department of Physics, Budapest University of Technology and MTA-BME Lend\"ulet Magneto-optical Spectroscopy Research Group, 1111 Budapest, Hungary}
\affiliation{RIKEN Center for Emergent Matter Science (CEMS), Wako 351-0198, Japan}
\author{Vladimir Hutanu}
\affiliation{Institute of Crystallography, RWTH Aachen University and J\"ulich Centre for Neutron Science (JCNS) at Heinz Maier-Leibnitz Zentrum (MLZ), 85747 Garching, Germany}\vspace*{3ex}

\date{\today}

\begin{abstract}
For a symmetry consistent theoretical description of the multiferroic phase of \bcgo a precise knowledge of its crystal structure is a prerequisite. In our previous synchrotron X-ray diffraction experiment on multiferroic Ba$_2$CoGe$_2$O$_7$ at room temperature we found forbidden reflections that favour the tetragonal-to-orthorhombic symmetry lowering of the titled compound. Here, we report the results of room-temperature single-crystal diffraction studies with both hot and cold neutrons to differentiate between the real symmetry lowering and multiple diffraction (the Renninger effect). A comparison of the experimental multiple diffraction patterns with simulated ones rules out the symmetry lowering. Thus, the structural model based on the tetragonal space group \PFTOm was selected to describe the Ba$_2$CoGe$_2$O$_7$ symmetry at room temperature. The precise structural parameters from neutron diffraction at 300\,K are presented and compared with the previous X-ray diffraction results.  
\end{abstract}

\maketitle                        

\section{Introduction}

Recently, multiferroic behavior and static magnetoelectric effects have been observed in many members of the melilite family, e.g., \ccso, \scso, \bmgo and \bcgo [see, e.g.,~\cite{prb.85.174106.2012} and references therein]. The dynamical magnetoelectric effect, also observed in these compounds, drastically changes the optical properties of multiferroics compared with conventional materials. For instance, a recent work of~\cite{nc.5.1.2014} discovered quadrochroism at the magnetoelectric spin excitations of multiferroic \ccso, \scso and \bcgo. 

Due to the lack of complete structural phase diagrams in the case of melilites, the high symmetry melilite phase is often used as a basis for theoretical calculations and experimental investigations [see, e.g., \cite{nc.5.1.2014}, \cite{prb.89.184419.2014}, \cite{np.8.734.2012}, \cite{apl.94.212904.2009} and references therein]. However, the melilite family shows a variety of structural phase transitions, including incommensurate phases, depending on temperature and chemical composition. The knowledge about the precise crystal structure is a very important input to understand magnetoelectric phenomena in multiferroics. For instance, in the work of~\cite{acra.67.264.2011} the main features of the magnetoelectric behaviour in \bcgo were predicted by symmetry considerations without referring to any specific microscopic mechanism. Thus, the exact structural model is often an essential starting point for further theoretical and experimental research.

The results of our previous synchrotron X-ray diffraction study on \bcgo below room temperature~(\onlinecite{prb.84.212101.2011}) are in favour of a symmetry lowering from the high-temperature tetragonal structure with space-group (SG) \PFTOm to the orthorhombic SG $Cmm2$. The assumption of the symmetry reduction was based on both the presence of reflections forbidden in \PFTOm~(\onlinecite{prb.84.212101.2011}) and the found magnetic space group (MSG) $Cm'm2'$~(\onlinecite{prb.86.104401.2012}). The MSG was then independently confirmed~(\onlinecite{prb.89.064403.2014}) by polarised neutron diffraction, namely the spherical neutron polarimetry technique. In contrast, the origin of the observed superstructure reflections in \bcgo was never cross-checked by other methods. An observation of the weak, but well noticeable intensities at the positions of forbidden reflections in reciprocal space for other melilite compounds, such as \ccso and \bmgo motivated us to re-investigate this effect more precisely.

In order to differentiate between real symmetry lowering and multiple diffraction (the Renninger effect), neutron diffraction with both short (hot neutrons) and long (cold neutrons) wavelengths was applied. Multiple diffraction patterns (the so-called $\psi$ scans) were simulated before the experiment in order to select the most appropriate wavelength and positions in reciprocal space. It was found, that the scattered intensity detected at the positions of forbidden reflections are entirely due to multiple diffraction. Thus, the crystal structure of \bcgo at room temperature (RT) can be perfectly described by the tetragonal SG \PFTOm. We present here the results of the model selection based on the multiple diffraction patterns as well as the complete structure refinement of \bcgo at RT using neutron diffraction.

\section{Experimental}

Single crystals of \bcgo were grown by the floating-zone technique and were well characterised in previous studies [see~\onlinecite{prb.89.064403.2014} and references therein]. The sample used for the neutron diffraction experiment has a cylindrical shape of approximately 5\,mm in height and about the same diameter.

Single-crystal neutron diffraction studies with both short and long wavelengths were performed on the two diffractometers HEiDi~(\onlinecite{nn.18.19.2007}) and MIRA~(\onlinecite{phb.397.150.2007}) at the FRM\,II reactor, Heinz-Maier-Leibnitz Zentrum (MLZ), Germany. On HEiDi, the wavelength $\lambda = 0.793$\,\AA\ was obtained from a Ge(422) monochromator with an Er-filter used to suppress the $\lambda/2$ contamination. A $^3$He point detector, optimised for neutrons with short wavelength, was used for the measurements of both $\psi$ scans and a full data collection at room temperature. On MIRA, the wavelength $\lambda = 4.488$\,\AA\ was obtained from a HOPG monochromator with Be-filter to suppress the $\lambda/2$ contamination. Both point and position sensitive detectors were used for the measurements of $\psi$ scans at room temperature.

On both instruments, an Eulerian cradle was used and the sample was oriented in a way to have the forbidden reflections of the ($h$00) type in the scattering plane when the $\chi$ angle in the conventional 4-cycle geometry was approx. 90\degree. This geometry allows an easy rotation around $\psi$ by 360\degree as it almost coincides with the conventional $\phi$ rotation. As the first step of the measurements, the orientation matrix was refined. Secondly, the conventional $\theta$, $\omega$, $\chi$ and $\phi$ values for any of the requested $\psi$ angles were calculated by the DIF4N~\footnote{Version of instrument control program DIF4~(\onlinecite{dif4.1992}) modified for neutron diffractometer HEiDi.} program. Then, an $\omega$ scan was performed for each $\psi$ value and the integrated intensity of the scan was assigned to that $\psi$ point. As a result, the integrated intensity versus $\psi$ curve was obtained and further compared with the calculations of possible multiple diffraction.     

The integrated intensities of the reflections collected with point detector were obtained using the DAVINCI program~(\onlinecite{davinci.sazonov.2015}). The multiple diffraction patterns were simulated using the UMWEG program~(\onlinecite{jac.40.185.2007}). The structural parameters of \bcgo were refined from the full data collection using the JANA2006 program~(\onlinecite{zfk.229.345.2014}). Experimental and refinement details are summarised in Table~\ref{t:SNDexperiment}.

\begin{table}[h]
\linespread{0.72}\normalsize
\caption{\label{t:SNDexperiment}Single-crystal neutron diffraction experimental and refinement details.}
\begin{ruledtabular}
\begin{tabular}{ll}
\emph{Crystal data}                       &                            					    \\
\colrule
Chemical formula                          & \bcgo                       					\\
$M_{\rm r}$                               & 590.8	                      					\\
Cell setting, space group                 & Tetragonal, \PFTOm		       					\\
$a$, $b$, $c$ (\AA)                       & 8.39(1), 8.39(1), 5.56(1) 						\\
$V$ (\AA$^3$)                             & 391.4	                      					\\
$Z$                                       & 2	                          					\\
$D_x$ (Mg\,m$^{-3}$)                      & 4.8877                       					\\
Radiation type                            & Constant wavelength 		 					\\
                                          & neutron diffraction 						 	\\
                                          & radiation, $\lambda=0.793$\,\AA					\\
$\mu$ (mm$^{-1}$)                         & 0.01	                       					\\
Temperature (K)                           & 300                          					\\
Crystal form, colour                      & Cylinder, blue		       						\\
Crystal size (mm)                         & $5\times 5\times 5$         					\\
                                          &                             					\\
\colrule
\emph{Data collection}                    &                             					\\
\colrule
Diffractometer                            & Four-circle diffractometer  					\\
Radiation source                          & Nuclear reactor             					\\
Monochromator                             & Ge~(422)	                  					\\
Temperature (K)                           & 300                          					\\
Data collection method                    & $\omega$ scans              					\\
$\theta_{max}$ ($^\circ$)                 & 49.45                        					\\
$[\sin\theta/\lambda]_{max}$ (\AA$^{-1}$) & 0.96                        					\\
Range of $h$, $k$, $l$                    & $0 \rightarrow h \rightarrow 14$ 				\\ 
                                          & $0 \rightarrow k \rightarrow 16$ 				\\ 
                                          & $-10 \rightarrow l \rightarrow 10$ 				\\ 
No. of measured reflections               & 1750                    						\\
No. of independent reflections            & 966	                    						\\
No. of independent reflections            & 952	                   							\\
~~~with $I > 3\sigma(I)$                  &                        							\\
$R_{\rm int}$                             & 0.019                   						\\
                                          &                         						\\
\colrule
\emph{Refinement}                         &                         						\\
\colrule
Refinement on                             & $F^2$                   						\\
$R[F^2>3\sigma(F^2)]$, $wR(F^2)$, $S$     & 0.019, 0.047, 2.09      						\\
No. of reflections                        & 952	                    						\\
No. of parameters                         & 35	                     						\\
Weighting scheme, $w$                     & $1/[\sigma^2(F_o^2)+0.0004F_o^4]$  				\\
Extinction correction                     & Isotropic,              						\\                                                         
                                          & Gaussian Type 1\footnote{According to \cite{acra.30.129.1974}}      						\\                                                         
Extinction coefficient                    & 0.137(2)              							\\
\end{tabular}
\end{ruledtabular}
\end{table}

\section{Results and Discussion}

\subsection{Room temperature crystal structure model}

In single-crystal neutron diffraction studies performed with short wavelengths ($\lambda \lesssim 1.2$\,\AA) on melilite compounds, we often observed weak reflections at the positions of reflections forbidden in the teragonal SG \PFTOm. The intensities of those reflections are usually larger than three times their standard deviations ($3\sigma$) and, in general, could be explained, by higher order wavelength contamination, multiple diffraction or symmetry lowering.

However, the forbidden reflections are found to be much stronger than possible contaminations by higher order wavelengths, which are effectively suppressed by the specific filters. Thus, at the HEiDi diffractometer with $\lambda=0.793$\,\AA, the $\lambda/2$ contribution is expected to be less than 0.5\,\%. In contrast, the experimentally measured intensities are much stronger and reach up to a few percents. Therefore, the higher order wavelength contamination is excluded from the list of the possible explanations of the observed superstructure reflections. 

In order to study an influence of the possible multiple diffraction contribution in \bcgo, a simulation of the $\psi$ scans was performed prior to the experiment. The instrumental parameters used in the simulation are the horizontal and vertical beam divergences ($\delta_H$, $\delta_V$) and the wavelength spread ($\Delta\lambda/\lambda$) of the incident beam. Those were estimated based on the geometry of the instrument. The \bcgo sample specific parameters, like the mosaic spread ($\nu$) and the mosaic-block radius ($r$) were set manually. Fractional atomic coordinates ($x$, $y$, $z$) and anisotropic atomic displacement parameters ($U_\mathrm{iso}$) for \bcgo were obtained from the structure refinement using the present single-crystal neutron diffraction data (see next section). 

\begin{figure} 
\includegraphics[width=0.955\columnwidth]{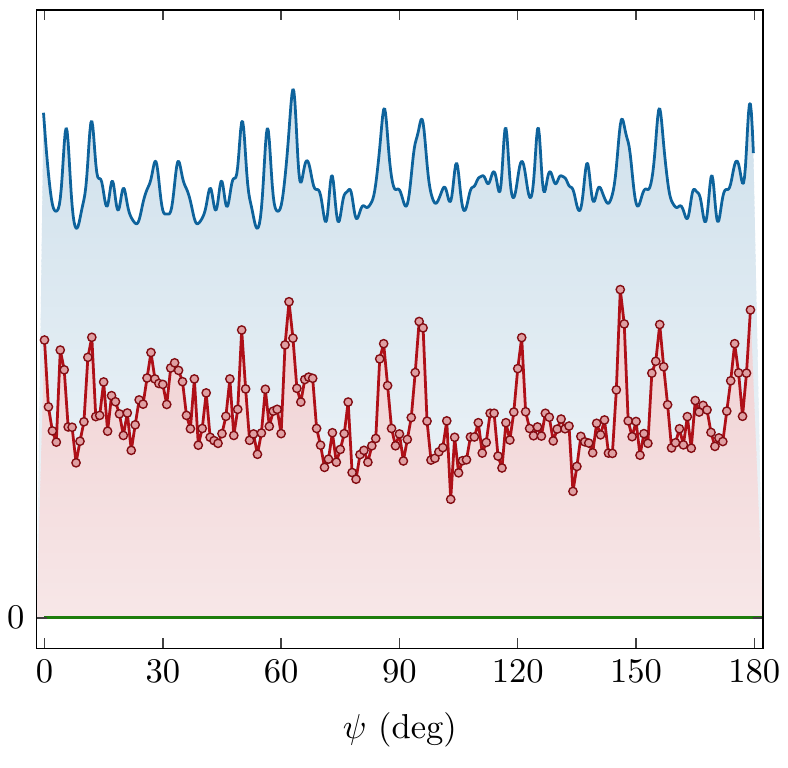}
\caption{\label{f:MultipleHeidi}Multiple diffraction pattern of the forbiden (300) reflection of \bcgo according to the calculations (top blue curve) and single-crystal neutron diffraction experiment (bottom red curve) with short wavelength $\lambda=0.793$\,\AA.}
\end{figure}

Figure~\ref{f:MultipleHeidi} shows the comparison of the simulated multiple diffraction pattern for the forbidden (300) reflection and the experimental data collected at HEiDi with $\lambda = 0.793$\,\AA. The following parameters were used in the calculations: $\Delta\lambda/\lambda \approx 0.01$, $\delta_H \approx 0.7^\circ$, $\delta_V \approx 1.6^\circ$, $r \approx 15$\,$\mu$m and $\nu \approx 0^\circ$. A good agreement between the calculated and experimental data indicates that both instrumental and sample specific parameters were reasonably selected. As can be seen from figure~\ref{f:MultipleHeidi}, the very broad peaks on the $\psi$ scan make it impossible to separate the case of the symmetry lowering from that of the multiple diffraction. There is no single point in the diffraction pattern without an overlap of the neighbouring reflections and the intensity never drops down to zero. As a result, the possible contribution caused by symmetry reduction could be well hidden by the multiple diffraction part. A comparison of the simulated and experimental intensities is not reliable and therefore could not be used to solve the problem.

\begin{figure} 
\includegraphics[width=0.955\columnwidth]{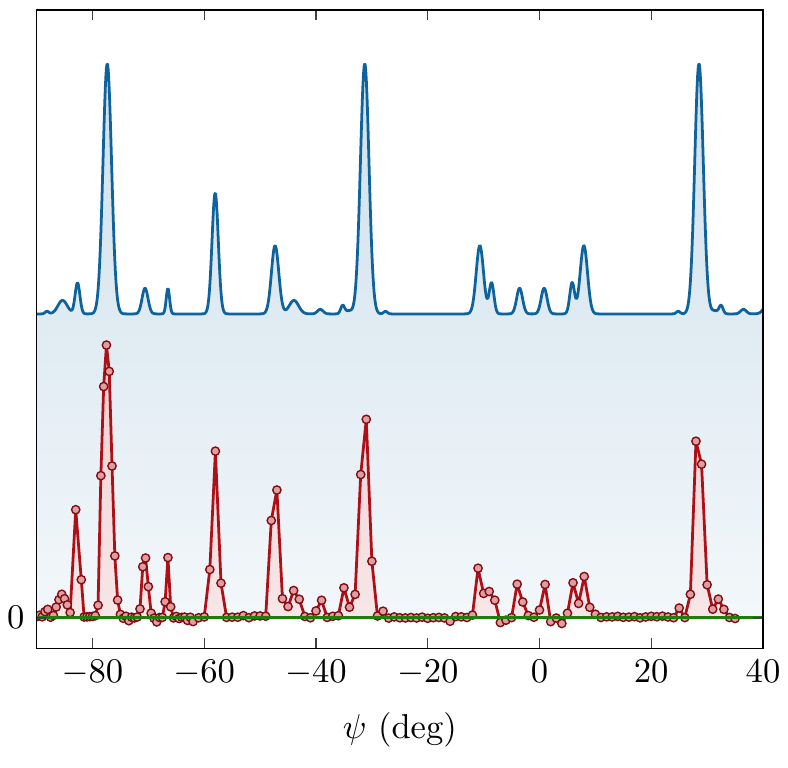}
\caption{\label{f:MultipleMira}Multiple diffraction pattern of the forbiden (100) reflection of \bcgo according to the calculations (top blue curve) and single-crystal neutron diffraction experiment (bottom red curve) with long wavelength $\lambda=4.488$\,\AA.}
\end{figure} 

In contrast to the short wavelength case, cold neutrons allow to overcome the problem with overlapping peaks in the $\psi$ scans. A change of just one single parameter, the wavelength $\lambda$, in the simulation process from $\lambda = 0.793$\,\AA\ to $\lambda = 4.488$\,\AA\ drastically modifies the entire multiple diffraction pattern. The number of peaks is reduced and they become well separated from each other: Regions of zero intensity appear on the simulated curve. An adjustment of the instrumental parameters allows to achieve a better agreement between the experimental data and calculations. Thus, in the simulations with cold neutrons, the following parameters were used: $\Delta\lambda/\lambda \approx 0.01$, $\delta_H \approx 0.5^\circ$, $\delta_V \approx 0.5^\circ$, $r \approx 15$\,$\mu$m and $\nu \approx 0^\circ$. We selected the forbidden (100) reflection for the measurement because the (300) reflection was not reachable with the long wavelength. The experimental values for (100) agree well with the calculation as can be seen in figure~\ref{f:MultipleMira}. We were able to reproduce the zero-intensity regions experimentally. Therefore, the observed intensity on the positions of forbidden reflections could be explained solely by multiple diffraction. This rules out the symmetry lowering scenario in \bcgo and supports the tetragonal SG \PFTOm to be the correct description of the true structure at room temperature. 

However, the question about the \bcgo symmetry in the antiferromagnetic state (below $T_\mathrm{N} \approx 5.7$\,K) remains open. In the magnetically ordered state at low temperatures, a separation between the symmetry lowering and multiple diffraction cases becomes much more difficult because of the allowed magnetic contribution to the positions of forbidden reflections on the $\psi$ scans (due to the extinction rules). In that case, the use of polarised neutron diffraction is preferable, which could allow to separate the nuclear part from the magnetic one. However, the geometrical limitations on the sample rotation and inclination caused by the necessity of using the cryostat significantly complicates the experimental setup. As a result, the method applied here is not easily applicable to unambiguously determine the crystal structure at low temperatures.

It should be noted, that in our previous synchrotron X-ray diffraction study on \bcgo below room temperature we also observed the forbidden reflections and interpreted them as a symmetry lowering~(\onlinecite{prb.84.212101.2011}). In that experiment, the test $\psi$ scan did not show a complete disappearance of the intensity at the positions of forbidden reflections. The instrumental parameters, such as, the beam divergence and wavelength spread are expected to be much smaller for the synchrotron X-ray case and as a result the peaks on the $\psi$ scan should be separated even at short wavelengths. However, the short wavelength ($\lambda \approx 0.7$\,\AA) combined with the possible large mosaicity of the sample used in that experiment could significantly increase the peak width and thus hide the zero-intensity regions.

\subsection{Room temperature structural details}

\begin{figure}
\includegraphics[width=1\columnwidth]{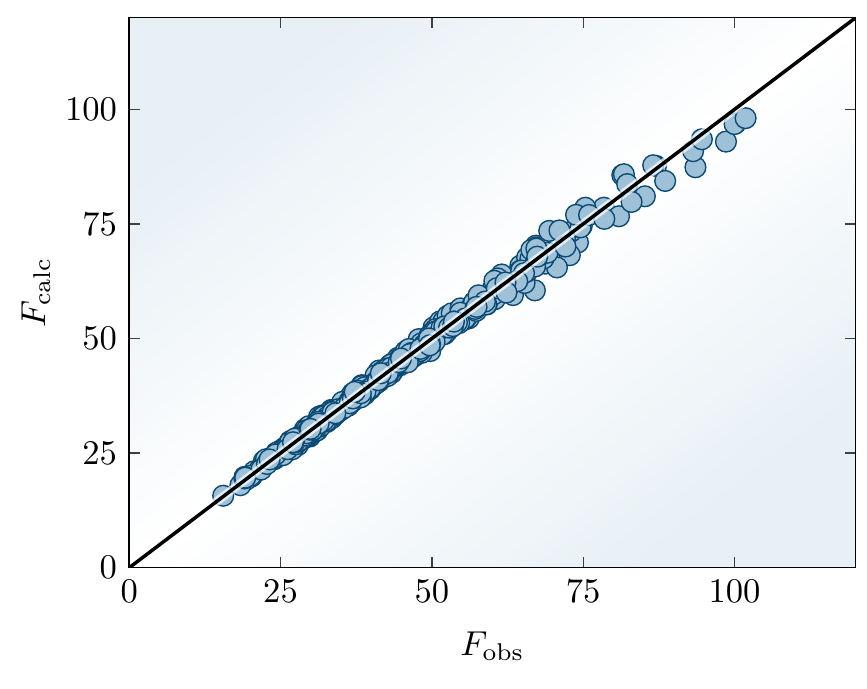}
\caption{\label{f:Fit}Quality of the \bcgo crystal structure refinement in SG \PFTOm according to the present single-crystal neutron diffraction data at 300\,K. The experimentally measured integrated intensities ($F^2_\mathrm{obs}$) are plotted against the calculated ones ($F^2_\mathrm{calc}$). The intensities of few selected Bragg reflections are marked.}
\end{figure}

In order to determine the precise structural parameters for \bcgo at room temperature we performed a refinement from our neutron diffraction data using the crystal structure model selected in the previous section (SG \PFTOm). The starting parameters were taken from the published structure determined by synchrotron X-ray diffraction at 293\,K~(\onlinecite{prb.84.212101.2011}). All atomic positions which are not restricted by symmetry were refined together with the anisotropic atomic displacements ($U_\mathrm{aniso}$), scale and extinction parameters. It should be noted here, that the intensity of strong reflections may be reduced by multiple diffraction, an effect which is not taken into account in the refinement calculations. The agreement between the experimental and calculated data is shown in Figure~\ref{f:Fit}. Table~\ref{t:RefinementXYZ} presents the refined atomic coordinates as well as the isotropic atomic displacement ($U_\mathrm{iso}$) parameters, while $U_\mathrm{aniso}$ are given in table~\ref{t:RefinementUaniso}. Full details of the refinement, including bond lengths and angles, are deposited in the crystallographic information file (CIF). 

\begin{table}
\caption{\label{t:RefinementXYZ}Fractional atomic coordinates ($x$, $y$, $z$) and isotropic atomic displacement parameters $U_\mathrm{iso}$ (\AA$^2$) for \bcgo refined in SG \PFTOm according to the present single-crystal neutron diffraction data at 300\,K.}
\begin{ruledtabular}
\begin{tabular}{lcrrrrr} 
Ion   & Wyckoff  & \multicolumn{1}{c}{$x$} 	& \multicolumn{1}{c}{$y$} 	& \multicolumn{1}{c}{$z$} 	& \multicolumn{1}{c}{$U_\mathrm{iso}$} 	\\ 
      & position &                         	&                         	&                         	& 										\\ 
\colrule
Ba    & $4e$     & 0.33475(6)				& 0.16525(6)				& 0.4926(2)					& 0.0072(1)                	        	\\
Co    & $2b$     & 0						& 0							& 0							& 0.0063(3)                     		\\
Ge    & $4e$     & 0.14052(4)				& 0.35948(4)				& 0.0398(1)					& 0.0053(1)                		       	\\
O1    & $2c$     & 0						& 0.5						& 0.1581(2)					& 0.0098(2)                     	   	\\
O2    & $4e$     & 0.13840(7)				& 0.36160(7)				& 0.7298(1)					& 0.0103(1)                     	  	\\
O3    & $8f$     & 0.07943(7)				& 0.18466(6)				& 0.1872(1)					& 0.0100(1)                     	  	\\
\end{tabular}
\end{ruledtabular}
\end{table}

A comparison of the room temperature structure from neutron and synchrotron X-ray diffraction shows a negligible difference in the positional parameters with an average value of less than $1\sigma$. The largest shift in the atomic positions of approx. 0.007\,\AA\ is found for $z_\mathrm{O2}$. 

\begin{table}
\caption{\label{t:RefinementUaniso}Anisotropic atomic displacement parameters $U_\mathrm{aniso}$ (\AA$^2$) for \bcgo refined in SG \PFTOm according to the present single-crystal neutron diffraction data at 300\,K.}
\begin{ruledtabular}
\begin{tabular}{lcccccc} 
Ion &\hspace*{4ex}& \multicolumn{1}{c}{$U_{11}$} &\hspace*{4ex}& \multicolumn{1}{c}{$U_{22}$} &\hspace*{4ex}& \multicolumn{1}{c}{$U_{33}$} \\ 
\colrule
Ba 	&&  0.0074(2)	 	&&  $U_{11}$ 	&&  0.0068(2) 	\\
Co 	&&  0.0056(3) 		&&  $U_{11}$ 	&&  0.0076(7) 	\\
Ge 	&&  0.0053(1) 		&&  $U_{11}$ 	&&  0.0053(2) 	\\
O1 	&&  0.0111(2) 		&&  $U_{11}$ 	&&  0.0071(4) 	\\
O2 	&&  0.0123(2) 		&&  $U_{11}$ 	&&  0.0065(2) 	\\
O3 	&&  0.0136(2) 		&&  0.0066(2) 	&&  0.0098(2) 	\\
\\
\colrule
Ion && \multicolumn{1}{c}{$U_{12}$} && \multicolumn{1}{c}{$U_{13}$} && \multicolumn{1}{c}{$U_{23}$} \\ 
\colrule
Ba 	&&  \,\,0.0024(2) 		&& -0.0005(1) 		&& -$U_{13}$		\\
Co 	&&  \,\,0				&&  \,\,0 			&&  \,\,0			\\
Ge 	&&  \,\,0.0006(1) 		&&  \,\,0.0001(1) 	&& -$U_{13}$		\\
O1 	&&  \,\,0.0061(3) 		&&  \,\,0 			&&  \,\,0			\\
O2 	&&  \,\,0.0023(2) 		&&  \,\,0.0006(1) 	&& -$U_{13}$		\\
O3 	&& -0.0030(1)			&& -0.0021(2) 		&&  \,\,0.0006(1)	\\
\end{tabular}
\end{ruledtabular}
\end{table}

\section{Conclusions}

In order to differentiate between a real symmetry lowering and multiple diffraction (the Renninger effect), neutron diffraction experiments with both hot and cold neutrons were performed. It was found that the scattered intensities detected at the positions of forbidden reflections are entirely due to the multiple diffraction. Thus, the crystal structure of \bcgo at room temperature can be described by the tetragonal SG \PFTOm without any symmetry lowering. The precise structural parameters of \bcgo at 300\,K are found to be in a good agreement with the previous synchrotron X-ray diffraction results.


\begin{acknowledgments}
We thank G. Heger for fruitful discussions. 
\end{acknowledgments}

\section*{References}

%

\end{document}